\documentstyle[11pt,newpasp,twoside]{article}
\def\edcomment#1{\iffalse\marginpar{\raggedright\sl#1\/}\else\relax\fi}
\reversemarginpar

\begin{document}
\title{Detection of Variable Circular Polarization in the SW Sex star V795 Herculis} 

\author{P. Rodr\'\i guez-Gil$^1$, J. Casares$^1$, I. G. Mart\'\i nez-Pais$^{1,2}$ and P. Hakala$^3$} 

\affil{$^1$Instituto de Astrof\'\i sica de Canarias, V\'\i a L\'actea, s/n, La Laguna, E-38200, Tenerife, Spain}
\affil{$^2$Departamento de Astrof\'\i sica, Universidad de La Laguna, Avda. Francisco S\'anchez, s/n, La Laguna, Tenerife, Spain}
\affil{$^3$Tuorla Observatory, University of Turku, Vaisalantie 20, Pikkio, FI-21500, Finland}

\begin{abstract}
We report the detection of modulated circular polarization in V795 Her. The degree of polarization increases with wavelength and is modulated with a period of 19.54 min, which is very close to the reported optical QPO period. The modulation has a peak-to-peak amplitude of 0.12~\% in the $U$-band. The estimated magnetic field intensity is in the range 2--7 MG.
\end{abstract}
\section{Introduction}
From the early 90's, many models have been proposed to explain the behaviour of the SW Sex stars. Currently, the main point of discussion is whether the SW Sex stars harbour magnetic white dwarfs or not. 
The idea of SW Sex stars being magnetic CVs was strenghtened after the discovery of variable circular polarization in LS Pegasi, modulated with a period of 29.6 min (Rodr\'\i guez-Gil et al. 2001). The authors introduced a new accretion scenario, in which a disc-overflown gas stream meets the white dwarf's magnetosphere at the corotation radius, where a shock above the disc is produced.
\section{The observations}
Simultaneous $UBVRI$ polarimetry was performed in 1998 March 2 and 5, with the {\sc turpol} photopolarimeter at the 2.56-m Nordic Optical Telescope, in La Palma. The object was observed for a total of 5.3 hours (2.04 orbital periods). The zero polarization star BD +33$\deg$2642 was also measured.
\section{Results}
The mean circular polarization levels, after subtracting the averaged levels of the comparison star are: $P_{\mathrm{c}}(U)=0.16 \pm 0.04~\%$, $P_{\mathrm{c}}(B)=0.16 \pm 0.05~\%$, $P_{\mathrm{c}}(V)=0.29 \pm 0.09~\%$, $P_{\mathrm{c}}(R)=0.24 \pm 0.07~\%$ and $P_{\mathrm{c}}(I)=0.41 \pm 0.13~\%$. This increasing trend with wavelength is reminiscent of cyclotron emission from accretion columns in Intermediate Polars (see e.g. West, Berriman, \& Schmidt 1987). The same trend is observed in the Intermediate Polars, PQ Gem (Potter et al. 1997) and BG CMi (West et al. 1987).\par
\begin{figure}[t]
\plotfiddle{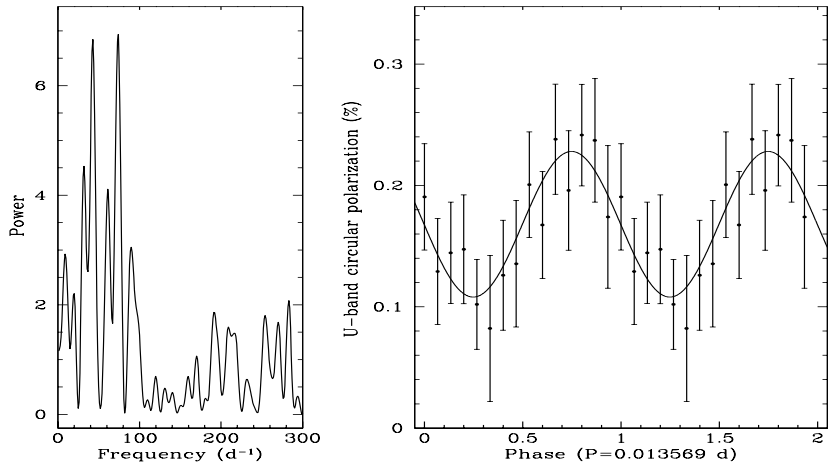}{6.5cm}{-90}{45}{30}{-180}{185}
\caption{{\it Left panel}: $U$-band Scargle periodogram. {\it Right panel}: $U$-band data folded on the 19.54-min period after averaging in 15 phase bins. The solid line is the best sine fit.}
\end{figure}
In the periodogram from the $U$-band data (see left panel of Fig.~1), two prominent peaks are located at $33.49 \pm 1.02$ min and $19.54 \pm 1.02$ min. We get a clear modulation when folding the $U$-band data on the latter period (see right panel of Fig.~1), obtaining a peak-to-peak amplitude of 0.12~\%. The modulation with this period is also evident in $B$-band and seems to vanish redwards. The 19.54-min period is consistent with the 19.3-min optical QPO reported by Patterson \& Skillman (1994). So, the 19.54-min period could be the spin period of a magnetic white dwarf. The polarization only takes positive values, a characteristic of low-inclination magnetic systems. Following the method proposed by Rodr\'\i guez-Gil et al. (2001) to estimate the primary magnetic field, we get a magnetosphere extending to $f = 0.55$ R$_{\mathrm{L_1}}$ (in LS Peg, $f = 0.52$ R$_{\mathrm{L_1}}$) and a field intensity of $B_1 \simeq$ 2--7 MG.

\end{document}